# Critical and maximally informative encoding between neural populations in the retina

David B. Kastner[1,3,5], Stephen A. Baccus[2], Tatyana O. Sharpee[3,4]

[1]Neuroscience Program, Stanford University School of Medicine, 299 Campus Drive W. Stanford, CA, 94305 USA. [2]Department of Neurobiology, Stanford University School of Medicine, 299 Campus Drive W., Stanford, CA, 94305 USA. [3]Computational Neurobiology Laboratory, the Salk Institute for Biological Studies, La Jolla, CA, 92037 USA. [4]Center for Theoretical Biological Physics and Department of Physics, University of California, San Diego, La Jolla, CA, USA. [5]Present address: Department of Psychiatry, University of California, San Francisco.

**Abstract:** Computation in the brain involves multiple types of neurons, yet the organizing principles for how these neurons work together remain unclear. Information theory has offered explanations for how different types of neurons can optimize the encoding of different stimulus features. However, recent experiments indicate that separate neuronal types exist that encode the same stimulus features, but do so with different thresholds. Here we show that the emergence of these types of neurons can be quantitatively described by the theory of transitions between different phases of matter. The two key parameters that control the separation of neurons into subclasses are the mean and standard deviation of noise levels among neurons in the population. The mean noise level plays the role of temperature in the classic theory of phase transitions, whereas the standard deviation is equivalent to pressure, in the case of liquid-gas transitions, or to magnetic field for magnetic transitions. Our results account for properties of two recently discovered types of salamander OFF retinal ganglion cells, as well as the absence of multiple types of ON cells. We further show that, across visual stimulus contrasts, retinal circuits continued to operate near the critical point whose quantitative characteristics matched those expected near a liquid-gas critical point and described by the nearest-neighbor Ising model in three dimensions. By operating near a critical point, neural circuits can optimize the trade-off between maximizing information transmission in a given environment and quickly adapting to a new environment.

## Introduction

Neural circuits use populations composed of multiple cell types to perform complex computations. Theoretical arguments, based upon the maximization of information transmitted about incoming stimuli, have proven successful in accounting for properties of single neurons[1-5] or populations of neurons encoding either one[6-10] or several different visual features[11-14]. However, recent experiments in the retina have discovered types of neurons that encode the same visual features but with different thresholds[15]. Here we sought to develop a framework to explain the existence of neuronal types distinguished by their nonlinear properties, taking into account not only noise in different neurons but also variation in the noise level between neural classes. We use the retina as a tractable system to study how neural responses in heterogeneous populations might be coordinated to efficiently encode complex sensory inputs.

In the salamander retina, populations of two types of Off cells encode nearly the same spatiotemporal visual feature, but separately tile the retina, indicating that they are distinct cell types[15]. These two types of neurons also maintain different thresholds across a range of contrasts, with differences in threshold between populations exceeding the variation in threshold within each type (Figure 1). Specifically, fast Off sensitizing cells maintain a lower threshold, encoding weaker signals, than fast Off adapting cells, which encode stronger signals. Notably, all types of Off cells split into such adapting and sensitizing subtypes, whereas the On types do not. These specific differences in the neural encoding between populations provide a particularly convenient model in which to analyze the factors that cause distinct neural populations to arise.

Given that the need for metabolic efficiency provides a strong constraint on the spike rate of neurons [16, 17], and presents a useful and necessary framework for understanding maximally informative solutions in neural circuits[3, 7, 9, 14], we sought to determine whether this coordination of thresholds among these cell types could be a consequence of maximizing the information these two populations can jointly provide about the input, given an energy constraint on the system.

The process of optimizing a population model to maximize a quantity such as information can be directly analogous to a physical system that attempts to find a minimal energy state, and a very mature literature exists to analyze, explain and unify the behavior in such physical systems[18, 19]. In our data, we found that the information maximization framework explained the separation of thresholds between adapting and sensitizing cells. We further found, by analyzing the model's optimal behavior, that a sharp discontinuity existed such that two populations should emerge below a specific level of noise, and two retinal populations were positioned near this discontinuity. This abrupt behavior corresponds to what in physics is known as a phase transition or a critical point. Systems near a phase transition exhibit a number of instabilities that may further increase the sensitivity of neural encoding to stimuli [18, 20-23]. Some signatures of phase transitions have been observed in neural circuits, including power law relationships between parameters that describe the susceptibility of the system to external perturbations and parameters that measure deviations of the system from the critical point, such as temperature and magnetization. The presence of such power law scaling relationships has indicated that circuit dynamics obey a regular structure across multiple time scales[20, 24-27], but the existence of this structure and its relevance to neuroscience remains a matter of debate[27, 28]. Furthermore a direct mapping onto one of the types of phase transitions that are known to occur in physical systems has remained elusive[25, 27].

Here, we quantitatively map the behavior of two retinal populations to the class of phase transitions that occur in magnetic systems as well as between liquids and gas near their critical points. We directly relate parameters from neural response functions, such as the noise level and the threshold, to the corresponding quantities in the physical systems, such as temperature and magnetization. This mapping makes it possible to take advantage of extensive research in physics[18, 19, 29] to quantitatively analyze and interpret [28] the large degree of sensitivity observed in populations of neural responses to external signals[20, 23-27]. By analyzing retinal responses in this context we conclude that retinal populations reside near a critical point where they are optimized both to maximize information and to respond quickly to changes in the environment.

## Results
**A model for quantifying information transmission in a population of multiple types of neurons**

Two populations of fast Off cells in the retina have very similar linear filtering properties but exhibit systematic differences in their response functions[15]. To measure these response functions we recorded the extracellular activity of ganglion cells in the presence of Gaussian white noise stimuli across a large range of different contrasts (Figure 1A and Supplementary Figure S1A). We calculated the linear-nonlinear (LN) model of the steady-state response at each contrast consisting of a linear temporal filter followed by a fixed nonlinear function (see Materials and Methods). The linear component of the model represents the feature to which, on average, the cell has the greatest sensitivity, while the nonlinearity of the model captures the dynamic range and response function of the cell in each contrast condition. This static nonlinearity describes the probability to observe a spike from a given neuron as a function of the strength of the stimulus projected onto the most relevant visual feature of the neuron. These two populations of fast Off cells have very similar linear filters, indicating they have the same preferred stimulus feature.

Previous studies have described methods that can be used to quantify how efficiently multiple classes of retinal neurons encode visual inputs[9, 11]. Specifically, the mutual information provided about natural scenes by the responses of multiple arrays of neurons as a whole can be approximated by a product of the information provided by the set of neurons representing the smallest repeating element within the array times the scaling factor that depends only on the array size. Thus, to find optimal parameters of neural responses functions, it is sufficient to consider the encoding of visual stimuli by the circuit's smallest repeating element. Here we focus on the encoding of visual stimuli as projected onto the set of visual features of fast Off cells, leaving the more general problem of encoding different input dimensions for future studies. For the problem at hand, the repeating element within the retina to consider consists of two neurons, one fast Off adapting and one fast Off sensitizing neuron. Because these neurons encode the same visual feature, we can consider the encoding of a one-dimensional signal, reflecting the output of the filtering of the visual stimulus by the relevant feature shared by these two types of neurons. We note that reducing two arrays of neurons to two representative neurons corresponds to the mean-field approximation in physics, which has proven successful in describing some of the most prominent aspects of emergent collective phenomena in physical systems[18].

The mutual information transmitted by the two neurons representing fast Off adapting and fast Off sensitizing cells can be increased or decreased by changing the parameters of the static nonlinearities of these two neurons. For each of the two neurons, we modeled the static nonlinearity as a sigmoid function with two parameters (Figure 1C and see Materials and Methods): µ, which is the input value that leads to a 50% spiking probability, and ν, reflecting the slope of the function, which is related to uncertainty in the neural response, with a large ν indicating a shallow slope with large noise (Supplementary Figure S1B). To understand the conditions under which maximal information transmission occurs for two such populations, we sought to optimize the position of the response functions for each neuron, while placing a constraint on the overall energy usage of the system by limiting the average rate, $\langle r \rangle$, across a population of cells (Supplementary Figure S1C).

We analyzed the optimal position for the two response functions that leads to the maximal amount of information provided about the input. First, we considered the case where the two response functions have the same slope. Here, the optimal coding strategy changes as a function of the steepness of the slope, and two distinct behaviors emerge (Figure 2A). In the regime with shallow slopes corresponding to large noise and ν, the maximally informative solution requires that the two neurons have identical thresholds, where the threshold difference, $m = \mu_2 - \mu_1$, is zero. This regime corresponds to redundant encoding, where maximal information occurs through combining two noisy identical measurements on the signal, $x$. This indicates that even redundant encoding[30] can be an optimal strategy for populations of neurons. This solution stops being optimal when ν decreases below a certain critical value $\nu_c$, where the optimal solution requires separate thresholds for the two neurons, and non-zero values of $m$ become optimal. This highlights the importance of taking into account noise, something shown to be key for optimal encoding with transcription factors[31], as well as in cases where spikes are summed across time[32] or a neural population[33]. Thus, the maximally informative solution for neural populations undergoes a sharp transition from one to two populations when the noise decreases below a critical value.

**Accounting for differences in the number of Off and On cell types**

This transition between two different encoding schemes offers an explanation to a previously perplexing result. In salamanders all Off populations are heterogeneous, splitting into adapting and sensitizing populations. However, the On population is homogeneous, having a lower threshold[34], and only displaying adaptation[15]. A possible explanation for the relative homogeneity of the On population could be that the noise in the On population is great enough that the optimal coding strategy would be redundant encoding. That predicts that the On response functions should have a shallower slope (larger ν) than the fast Off cells. We found this to be the case (Figure 2B,C). Not only do On cells have shallower slopes than Off cells as previously reported[35], but their average slope lies above the critical point, placing them within the regime where redundant encoding is the optimal solution.

**Fast Off populations optimally space their response functions**

We then examined the information maximization framework by considering the case where the two response functions have unequal slopes. Here, the theory predicts that neurons with steeper slopes should have their thresholds closer to the mean (Figure 3A). This prediction was confirmed by experimental data (Figure 3B). Across the full range of contrasts, sensitizing cells maintained a steeper slope than adapting cells. As predicted by the model, sensitizing cells have the lower threshold (Figure 1B), and have less noise, as manifested by their steeper slope.

Up until this point we have used this theoretical framework to understand general features of the retinal data; however, it also has the capacity to determine the optimal threshold difference for individual pairs of fast Off ganglion cells. By using the slopes and rate constraint from the data we determined how information transmission changed as the threshold difference between the response functions varied (Figure 3C). This theoretical framework not only accounted for general features of the data, but it also accurately predicted the separation of thresholds among individual pairs of simultaneously recorded adapting and sensitizing cells (Figure 3C). For all contrasts, the threshold difference of the data was very close to the threshold difference that maximizes the information about the input. On average, fast Off cells had a threshold difference that provided >97% of the maximal amount of information across a large range of contrast distributions (Figure 3D). Thus, these populations of neurons maintained the optimal position of their response functions even when the average spiking probability for the adapting and sensitizing cells increases with increasing contrast (Figure 1B and Figure S1).

**Maximally informative solutions undergo a second-order phase transition**

The behavior depicted in Figure 2A, where the optimization function—here maximizing information—transforms from having a single optimal state to having two divergent optima, is one of the signatures of a second-order phase transition[18]. Positioning neurons near a phase transition bears with it implications about the dynamics and sensitivity of neural circuits, and these implications depend upon the type of phase transition[18]. To rigorously establish whether the transition that we find in our model corresponds to one of the known classes of phase transitions in physics we have to find the appropriate correspondences between the key parameters that govern phase transitions in physics with the parameters used in our model. The theory of phase transitions draws its power from its ability to encompass diverse types of complex physical systems, mapping equivalent parameters from different physical systems onto each other. Even though some of the correspondences are not immediately intuitive, such as why density near the liquid-vapor critical point should correspond to magnetization in Ising magnetic systems, the theory provides a framework to explain why both of these quantities follow identical power law dependencies with respect to temperature[18]. As the first correspondence, we note that the observed states of matter come about through minimizing the free energy; in our solutions this corresponds to maximizing the information. Second, in physics, transitions occur with respect to temperature. In our case it occurs as a function of the average slope, $v = \frac{v_2 + v_1}{2}$, of the two response functions, which describes the average noise in the neural responses. Third, in the Ising model, magnetization spontaneously appears below the critical temperature; here, the corresponding quantity is the threshold difference, $m = \mu_2 - \mu_1$, between the thresholds of the optimal response functions, which takes non-zero values below the critical noise level. Finally, to find the quantity analogous to the applied magnetic field, we note that an applied magnetic field induces non-zero magnetization even above the critical temperature. In our case, we find that a difference in the noise in the two response functions, $h = v_2 - v_1$, induces a non-zero optimal threshold difference between thresholds of the response functions over a broad range of noise levels (Figure 3A). This suggests that the difference in the noise in the two response functions, $h$, or more generally the standard deviation of $v$ values across the neural ensemble, is analogous to an applied magnetic field. We have also verified this correspondence quantitatively by showing that $h$, computed as the derivative of information with respect to magnetization, is linear with respect to $v_2 - v_1$ (Figure S3). In Figure 3A, the curves for different $\langle r \rangle$ values are plotted for the same average noise value $v$. Because the critical noise value $v_c$ depends on the average firing rate $\langle r \rangle$ of the two neurons, the curves for different $\langle r \rangle$ are effectively placed at different distances from their respective critical points. This accounts for their spread.

Using these identified correspondences we can now determine the type of phase transition that occurs in our model with decreasing noise level. The properties of the phase transition can be characterized by examining specific discontinuities—or singularities—present in the derivatives of the optimization function, which in our model maximizes information. Specifically, in most cases a phase transition is defined as a second-order transition when the singularities appear in the second derivatives of the optimization function. First, we examined the second derivative of information with respect to noise (Figure 4A). This quantity is analogous to the specific heat, $C = \partial^2 I / \partial v^2$. We observe that $C$ is largely independent from v on each side of the transition with a sudden drop across the critical point. This is precisely the singular behavior expected based upon mean-field calculations for magnetization in magnetic systems, with smaller values observed above the critical temperature [19]. Second, we found that the second derivative of the information with respect to $h$, $\chi = \partial^2 I / \partial h^2$, with $h$ being the difference in slope of the response functions, displays a singularity at the critical point (Figure 4B). This function $\chi$ is analogous to the magnetic susceptibility in magnetic transitions, which is the second derivative of the energy with respect to an applied magnetic field[18], and is sometimes interpreted as describing the system's sensitivity to external perturbations[22]. Mean-field calculations indicate that this quantity should decay with an exponent of -1 as a function of temperate difference from its critical value[18]. This matches the estimated exponent of -0.93 in our model (the difference from -1 reflect imprecision of numerical simulations). Thus, the transition we observe in neural circuits quantitatively matches behavior of the Ising model near its critical point.

At the critical point when $v=v_c$, the difference $m$ between thresholds is zero and the difference in the noise of the two response functions $h$ is also zero. The distance of the system from the critical point can thus be quantified by measuring how much the difference in thresholds $m$ and the difference in the noise in the two response functions $h$ differ from zero, as well as by the deviation of average noise from its critical value $v - v_c$. Near the critical point, these three quantities are not independent, but rather scale as power-law functions of each other. Specifically, the mean-field Ising model calculation predicts that $m \propto |v - v_c|^\beta$ and $m \propto h^{1/\delta}$ with β=1/2 and δ=3, respectively [18]. We found that both of these relationships held true for our model of neural populations: with exponents β=0.47 (Figure 4C), and 1/δ=0.34 (Figure 4D) both of which closely matched their theoretical values of 0.5 and 0.33[18], respectively. Therefore, by all metrics a system that maximizes information transmission in two populations falls into the class of models described by the Ising model of magnetism, which is called the Ising model universality class.

**Fast Off populations remain poised at the critical point**

To examine where retinal circuits are positioned relative to the critical point, we measured where the threshold and slopes of each cell pair lied relative to the critical noise value $v_c$. However, we needed to account for the fact that the average spike rate $\langle r \rangle$ differs for each pair of adapting/sensitizing neurons, and certain parameters such as $v_c$ vary with $\langle r \rangle$. Fortunately, the dependence of $v_c$ on $\langle r \rangle$ is stereotypic and smooth (Figure S2A), and could, therefore, be normalized for each cell pair. In addition, in the equation that relates the threshold difference $m$ to the critical noise value $m = A|v - v_c|^\beta$, (Figure 4C) the dependence of the coefficient $A$ on the average rate is also smooth (Figure S2A). Finally, the scaling exponents β and 1/δ do not depend on $\langle r \rangle$ (Supplementary Figure S2B). This makes it possible to transform the data into normalized coordinates where variables $m$, $h$, and $v_c$ do not depend on $\langle r \rangle$ (Figure 5A). In these normalized coordinates, we can view all of the data from multiple pairs of cells relative to their respective critical points (Figure 5B). We find that for all pairs of cells across the full range of contrasts the fast Off ganglion cells reside below the critical noise value, in the regime where it is optimal to split the encoding between response functions with two different thresholds. Thus, although the position of the critical point changes with mean spike rate, fast-Off adapting and sensitizing cells maintain their response functions to stay below the critical point. We also note that while all of the results so far have been obtained for Gaussian signals, none of the results change in a noticeable way when more natural, non-Gaussian distributions are used as inputs (Figure S4). This robustness reflects the universality properties of systems near phase transition were many microscopic details become irrelevant.

**Scaling exponents in the retina match the Ising model universality class**

Remarkably, we find that even the deviations between the mean-field theory predictions and the experimental measurements in the retina matched the deviations observed in experiments on physical systems. Experimental measurements for the exponent $\beta$ fall within a narrow range from $0.316 - 0.34$ for all physical systems within the Ising universality class, including liquid-gas transitions in various substances, as well as ferromagnetic and anti-ferromagnetic transitions[18]. Fitting our experimental data (Figure 5B), we find a value, $\beta_{retina} = 0.39 \pm 0.12$ (Table 1). This value is consistent with the experimental observations in physical systems described by the three-dimensional nearest-neighbor Ising model[36] and deviates from values expected for Ising models with nearest-neighbor interactions in both two-dimensional and four-dimensional cases (Table S1). Similarly, experimental values for the exponent $1/\delta$ in systems from the Ising model universality class are shifted from the mean-field prediction of ⅓ to $0.204 - 0.217$[18]. Our data shifts in the same direction, with a value $1/\delta_{retina} = 0.15 \pm 0.08$. This value is also consistent with the nearest-neighbor Ising model in three dimensions. Thus, the direction and magnitude of deviations that we observe here in the scaling exponents suggest that they have the same origins as the deviations observed in the physical systems from their mean-field values. This quantitative agreement in the way physical systems deviate from the mean field predictions and the way the retina deviates from our mean field theory derived model further supports the use of this simplifying framework in characterizing the critical and maximally informative behavior of the retina.

An important aspect of adaptive neural systems is that they adjust quickly to changing environments, thus the speed of adaptation provides an additional factor that should influence response properties beyond a simple maximization of information transmission in one steady environment. The dynamics of critical systems have been thoroughly studied, and it is known that near a critical point, such systems exhibit a sharp transition between regions of fast and slow dynamics when their parameters change[18]. Because it is known that the threshold difference between curves is adjusted during adaptation[15], we analyzed whether the chosen steady state solution lied in a region that would allow fast dynamics should the environment suddenly change. We examined the positioning of the steady state retinal data relative to the theoretical curve dividing regions of fast and slow dynamics, the so-called 'spinodal line'[18]. We found that for nearly all of the data points, rather than being exactly at the optimal solution—which would maximize information but would be in a region of slow dynamics—the data lied near the spinodal line, meaning that they were in a region where a small change in the difference between thresholds yields large increases in information transmitted (Figure 5B). Because the steady state value of the response curves lied in a region that allowed fast dynamics, this indicates that the specific solution observed optimizes a tradeoff between maximizing information in a given environment, and adapting quickly to a new environment.

**Discussion**

In this work we have shown that the information maximization framework accounts for a number of properties in the retina and raises the possibility that cell types elsewhere in the brain are established and operate according to similar principles. First, information maximization accounted for both the lower thresholds (Figure 1B) and the steeper tuning functions (Figure 3B) of sensitizing cells compared to adapting cells. These results expand upon a set of previous studies where information maximization accounted for the diversification and coordination within a set of relevant input features in a neuronal population[4, 9-11, 13, 14, 37-41] or in multiple features affecting the responses of single neurons as predicted theoretically[42] and observed experimentally[43, 44]. Second, when considering neurons tuned to the same kind of input feature, we find that maximally informative solutions undergo a bifurcation when the steepness of neural response function exceeds a certain value (Figure 2A). Such a transformation in the relative parameters of neural nonlinearities augments the wide range of possible changes in the preferred stimulus features that are known to occur upon a changing noise level[45, 46] and/or the size of the neural population[47-49]. For example, decreasing the noise level can make the non-Gaussian characteristics in the input ensemble more apparent and this causes a shift in the relevant features from center-surround to oriented features[46]. Such transitions likely do not correspond to a phase transition, or at least at present the procedure for mapping such a transition onto a phase transition is not clear.

In our study, the maximally informative solutions were analyzed for a fixed average firing rate (these values were matched to experimental observations). Without this constraint, instead of two kinds of Off neurons, one obtains an Off neuron and an On neuron[12]. These bifurcations represent a robust phenomenon, because they persist even when the combinatorial representation of neural responses is simplified to a pooling rule where the identities of single neurons in the population are ignored[33], for bell-shaped tuning curves optimized to maximize Fisher information[50]. When more than two neurons are tasked with encoding the same one-dimensional signal, one can expect to find a series of bifurcations that progressively split the population into sub-groups as neural noise decreases[33]. Simultaneous splitting into three or more sub-groups might be possible, although likely to be rare, requiring special symmetry constrains involving the precise value for the average spike rate and number of neurons in the population. Elucidating these effects represents a promising direction for future studies.

The fact that the observed steepness of the response function of On cells and Off cells fell on different sides from the critical value (Figure 2B,C) represents a strong quantitative test of the theory. This match was obtained without any adjustable parameters, because the average firing rate extracted from experimental values uniquely determines the critical point. The error-bars for the critical point value in Figure 2C reflect the variation in the spiking rate between the recorded pairs of neurons. The information maximization framework thus explained both the relative homogeneity of On cells, which do not split into separate classes, and the presence of two classes of Off neurons. If the response functions of Off neurons were even steeper, one would potentially expect to find more classes of Off neurons. Thus, whether or not heterogeneity among neural response functions[51] improves information maximization can depend on the average steepness of these response functions.

A third feature of retinal processing highlighted by comparison with maximally informative solutions likely reflects the need to quickly adapt to changes in the input distribution. In a natural sensory environment, changes in contrast occur often and unpredictably[52]. To accommodate these changes in contrast, the slope of the response functions must adapt accordingly[2, 53-56], necessitating an accompanying slow change in the threshold differences between the response functions to maintain optimal encoding. The fact that this process cannot and does not occur instantaneously[54, 57, 58], leads to the following trade-off. In a stationary regime, which corresponds to the analyses that we have carried out, neurons with sharper tuning functions (smaller $v$) provide a greater amount of information (Figure 2A). However, more narrow tuning functions, corresponding to a lower temperature, require more time to reach the optimal state. This phenomenon, known in physics as a critical slowing down[59], occurs in the neural context because stimulus values that fall within the saturating region of the response function cannot be measured accurately, and thus cannot trigger adaptation. Therefore, by utilizing values of $v$ near $v_c$, which provide less absolute information (Figure 2A), the retina could be choosing an optimal strategy that it can reach quickly, rather than transmitting the largest possible amount of information at the expense of long adaptation times. These arguments are supported by experimental measurements in the retina where the observed values of threshold difference *m* achieve >97% of the maximum information while being invariably smaller than the optimal separation (Figure 5B). Tellingly, the data points lie near the so-called spinodal line that delineates the regions between fast and slow dynamics near a critical point (Figure 5B). The dynamics necessary to increase *m* from the suboptimal point of zero (redundant encoding) to the spinodal line only requires infinitesimal perturbations, while the dynamics of going from the spinodal line to the optimal line is expected to be slow, requiring large fluctuations[18]. The clustering of points near the spinodal line highlights a potential tradeoff between the need for optimality and the dynamics of adaptation.

We have drawn upon a well-established theory developed in physics to understand complex systems with many interactive degrees of freedom, in order to explain why neurons in the brain form new classes in order to maximize their information transmission. The mapping between maximally informative solutions and the theory of phase transitions in physics is both conceptual as well as quantitative (Table 1). Some of the connections make intuitive sense—noise in neural responses corresponds to temperature in physics—while other connections are more involved but respect the general properties of the information function, e.g. that it is an even function of *m* (Figure 2A) and an odd function with respect to $h=v_2-v_1$ (Figure 3A).

Second-order phase transitions often separate states that have different 'symmetry properties,' meaning that on one side of the transition, states remain constant when certain parameters of the system are changed, while states on the other side of the phase transition would be affected by the same change in parameters. For example, maximally informative solutions, within our model, exhibit different symmetry properties on different sides of the phase transition (two distinct cell types vs. one). Specifically, solutions in the large noise regime (one cell type) display greater symmetry because they are not affected by an exchange in neural indices since the thresholds are the same for the two neurons. For noise levels smaller than critical, $v<v_c$, the optimal solutions do not have this symmetry, because one of the neurons has a higher threshold, and this transformation would correspond to a change in the threshold difference from *m* to -*m*. The system has to choose a positive or a negative value by assigning one neuron to the adapting class and the other neuron to the sensitizing class. This process is directly analogous to magnetic systems where the system has to choose between two alternatives – "up" or "down" magnetization states – below the critical temperature, and thus these symmetry properties coincide with those of the Ising model. Thus, the arguments based on symmetry also indicate that the difference in thresholds, $m = \mu_2 - \mu_1$, is the neural quantity analogous to magnetization. Other quantities that have been proposed to correspond to magnetization include the mean spike rate[22] and the balance between excitation and inhibition[20]. While these quantities are potentially related to each other, and to the difference in thresholds *m* we describe here, in nonlinear ways, they do not produce a mapping onto one of the known type of phase transitions [22]. In addition, for these other neural quantities there is not an obvious change in symmetry on different sides of the transition we identify here.

One of the key properties of the theory of phase transitions is that it identifies which parameters are universal, i.e. independent of microscopic detail of any given system. The scaling exponents that characterize discontinuous behavior observed near a phase transition comprise such universal parameters. In contrast, other parameters, such as the critical temperature as well as the coefficients of proportionality in the scaling relationships, e.g. the constant *A* in $m = A|v-v_c|^\beta$, depend upon microscopic details and vary between systems. This separation into universal and non-universal parameters also held true in our model of neural circuits. In the case of retinal circuits, models with different average spike rate $\langle r \rangle$ yield the same value of scaling exponents *β* and 1/*δ*, whereas values of the critical noise and proportionality coefficients vary with $\langle r \rangle$ (Supplementary Figure S2A,B).

Similarly, the universal exponents as well as qualitative features of the transitions were unchanged when we considered non-Gaussian inputs (Figure S3).

Overall, the set of correspondences described in Table 1 between the physical quantities and their counterparts in neural coding have yielded a perfect, quantitative match in the type of singularities that are observed in the two fields of science. The match between the power law exponents that characterize the behavior of the maximally informative solutions near the critical point and those from mean-field theory calculations in physics (Figure 4C, D, and Table 1) can be summarized by stating that in both systems the behavior of the optimization function near the critical point can be approximated as

$$I \propto A(v - v_c)m^2 + Bm^4 + Chm, \qquad (1)$$

where A, B, and C are constants. This expression corresponds to the Landau theory of phase transitions[19]. The simple properties of this expansion formalize the argument that a neural circuit will be robust to changes that do not affect the control parameters, which consist of the mean $v$ and standard deviation $h$ of noise levels across the population. These control parameters in turn determine the optimal standard deviation of thresholds $m$.

It should be emphasized that our theoretical derivations were obtained to highlight differences in the response functions of neurons from different subpopulations, such as Off adapting and Off sensitizing cells, while ignoring differences in the response functions within each population. This corresponds to the mean-field approximation treatment of phase transitions in physics. It is well-known that mean-field theory summarized by the expansion in Eq. (1) can capture qualitative features of system behavior near the critical point, but its predictions for scaling exponents deviate in a systematic way from experimental measurements. In physical systems, these discrepancies have been resolved through the development of the renormalization group theory that builds upon on mean-field approximation but then takes into account the fluctuations in control parameters across the array. Notably, our measurements of scaling exponents in the retina matched experimental measurements in physical systems that correspond to the three-dimensional Ising model (Figure 5B and Table 1). This fact suggests that the deviations between measurements and mean-field theory predictions that are currently available can be resolved using the techniques from renormalization group theory to take into account small differences in response parameters across the retinal array.

The precision with which scaling exponents could be computed from retinal data was sufficient to rule out matches to nearest-neighbors Ising models of dimensions smaller or larger than three (Table S1). The values for the four-dimensional model coincide with values obtained using mean field theory and with values assuming infinite range of interactions[36]. One may wonder how the match to three-dimensional exponent can be consistent with retinal ganglion cells arranged in a two-dimensional (2D) array. It turns out that, in terms of critical exponents, an Ising model based on nearest neighbor interactions is equivalent to a lower dimensional model where interactions extent beyond nearest neighbors[60]. How fast interactions should increase in two dimensions depends on the dimensionality of the nearest-neighbor Ising model that we would like to match[60]. Given that retinal receptive fields are centered on a 2D lattice, our finding that critical exponents match the 3D Ising model implies that the effective interaction strength between neurons decreases as $r^{-3.309}$, with $r$ being the distance between receptive field center positions. Thus, the perspective from the theory of phase transitions yields very specific predictions that can be tested in future studies with large-scale multi-electrode recordings.

It is worth mentioning that, unlike in the physical systems where temperature is under complete experimental control, here we could not directly adjust $v$, which limits our ability to precisely estimate the scaling exponents of our system from experimental data. For the same reason, we cannot directly measure a quantity analogous to the specific heat, $C = \partial^2 I / \partial v^2$, which in our model relates the second derivative of the information with respect to noise (Figure 4A). Importantly, while many quantities exhibit singularities and power-law scaling relationships with each other (Figure 4), only two scaling exponents are independent and determine the type of the phase transition[18]. Thus, the match that we obtain between the neural and physical systems for the two exponents $\beta$ and $\delta$ is sufficient to identify the universality class of a phase transition. These observations might also help interpret the increase in specific heat curves observed in multi-electrode recordings[24]. In physical systems where the phase transition falls into the Ising model universality class, the specific heat curve exhibits a cusp-like singularity where the function is actually continuous across the transition, but its first derivative experiences a discontinuous jump. While this type of singularity is different from a discontinuity predicted by the mean field theory (Figure 4A), it is consistent with observations from large retinal populations[24]. This provides another example of a deviation observed in the retina that follow the deviations expected from physical systems when compared to the predictions from the mean-field theory.

In sum, it is fitting that a theory developed in physics to tackle the case of complex systems with many interactive degrees of freedom can also offer insights into the function of neural circuits. Perhaps broader application of the ideas described here could explain the existence of other classes of cells throughout the brain.

**Materials and Methods**
**Ethics Statement.** Experimental data were collected using procedures approved by the Institutional Animal Care and Use Committee of Stanford University, and in accordance with National Institutes of Health guidelines. Experimental and surgical procedures have been described previously[15].
**Experimental preparation.** We recorded from retinal ganglion cells of larval tiger salamanders using an array of 60 electrodes (Multichannel Systems) as previously described[15]. A video monitor projected the visual stimuli at 30 Hz controlled by Matlab (Mathworks), using Psychophysics Toolbox[61, 62]. Stimuli were uniform field with a constant mean intensity, $M$, of 10 mW/m$^2$ and were drawn from a Gaussian distribution. Contrast is defined as $\sigma = W/M$, where $W$ is the standard deviation of the intensity distribution. Neurons were probed with 9 different contrasts distributions from 12 – 36% in 3 % intervals. The contrasts were randomly interleaved and repeated. Each contrast was presented, in total, for $\geq 600$s. For the calculation of the response functions, the first 10 seconds of data in each contrast was not used to allow for a better estimation of the steady state.
**Linear-Nonlinear models.** LN models consisted of the light intensity passed through a linear temporal filter, which describes the average response to a brief flash of light, followed by a static nonlinearity, which describes the threshold and sensitivity of the cell. To compute the model, the stimulus, $s(t)$, was convolved with a linear temporal filter, $F(t)$, which was computed as the time reverse of the spike triggered average stimulus, such that

$$g(t) = \int F(t-\tau)s(\tau) \qquad (2)$$

A static nonlinearity, $N(g)$, was computed by comparing all values of the firing rate, $r(t)$, with $g(t)$, and then computing the average value of $r(t)$ over bins of $g(t)$. The filter, $F(t)$, was normalized in amplitude such that it did not amplify the stimulus, i.e. the variance of $s$ and $g$ were equal[53]. Thus, the linear filter contained only relative temporal sensitivity, and the nonlinearity represented the overall sensitivity of the transformation.

The nonlinearity was fit using a logistic function:

$$P(spike\,|\,x) = \frac{1}{1+\exp\left(-\dfrac{x-\mu}{\nu}\right)} \, . \qquad (3)$$

This sigmoid function had two parameters, $\mu$, which is the $x$ value that leads to a 50% spiking probability, and the slope $\nu$ of the function (Figure 1C). The slope, $\nu$, of the sigmoid describes noise present in the system (Figure S1B), because the probability of the neural response is not simply 0 or 1 for a given input. With an infinitely steep slope ($\nu$=0), the sigmoid turns into a step function with no uncertainty in the neural response for a given input. This functional form is advantageous because it matches well the input functions of single neurons[3], and it represents a minimal function consistent with the constraints on the mean firing rate and mean stimulus, $x$, given a spike[63].

For Figure 3B,C, the stimulus was composed of independent 50 µm bars, each with a contrast distribution of 35 %. The fact that the stimulus was spatial made the exact contrast experienced by the cell unknown. However, the contrast for the normalization was estimated by comparing the values of the slopes to those recorded in uniform contrast (Figure 3B).

**Information maximization.** The two response functions from Eq. (3) also define the probability, $p(r_i | x)$, of observing one of the four possible responses of the two neurons given an input $x$. The four response patterns $r_i$ correspond to the presence or absence of a spike from each of the two neurons. This treats the neurons as conditionally independent without significant correlations in their responses for a given input, which is a good first approximation for the fast Off adapting and sensitizing cells[15]. From the response functions $p(r_i | x)$ one can also compute the average probability of observing $r_i$ by multiplying the response functions by the input distribution and averaging with respect to $p(x)$. This is all that is necessary to calculate the mutual information, which is the difference between the total entropy and the noise entropy.

The total entropy, $H(r)$, is computed by:
$$H(r) = -\sum p(r_i) \log_2 p(r_i). \tag{4}$$

The noise entropy, $H(r|x)$, is computed by:
$$H(r|x) = -\int p(x) \sum p(r_i|x) \log_2 p(r_i|x) dx. \tag{5}$$

For both entropy calculations the sum is taken over the four response probabilities. The sigmoid fits to the data were normalized by their maximal firing rate to ensure that they ranged from 0 to 1.

**Spinodal line, and model normalization.** The spinodal line was extracted from the curves that related the information to the threshold difference between the two response functions (Figure 3C), and is defined as the point between the minimum and maximum of the information where the derivative of information with respect to the threshold difference, $m$, is maximal. The spinodal points were determined for each pair of fast Off adapting and sensitizing cells at each contrast.

**Acknowledgments:** We thank Pablo D. Jadzinsky, John Berkowitz, Johnatan Aljadeff, and James E. Fitzgerald for helpful discussions, Oleg Barabash for help with numerical simulations, and Sreekanth H. Chalasani for comments on the manuscript.

**Figures and Supporting Information**

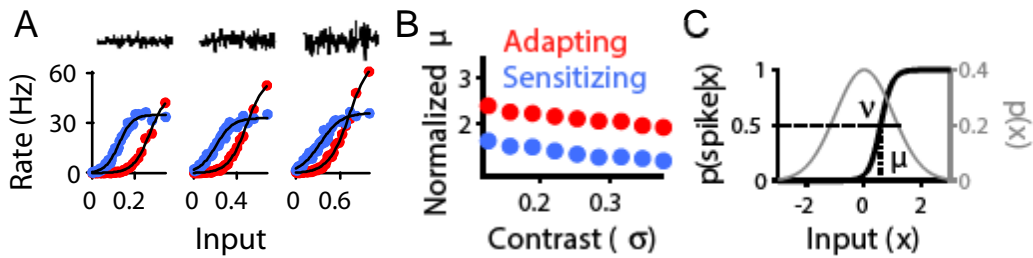

**Figure 1**. **Coordination between the thresholds of adapting and sensitizing fast Off cells.** (**A**) Steady state nonlinearities for adapting and sensitizing cells recorded simultaneously at three different levels of temporal contrast: 12 (left), 24 (middle), and 36 % (right). Example stimuli are shown in the top row. Black lines are the sigmoid fits to the data. (**B**) Adapting cells maintain higher thresholds than sensitizing cells across a range of contrasts. Symbols show average midpoint of the sigmoid, $\mu$,

normalized by contrast, σ, for all sensitizing (n = 11) and adapting (n = 36) cells at each contrast. Error values, s.e.m., are smaller than symbols. (**C**) Characterizing neural responses with logistic functions. Parameters of the response function are measured in units of the standard deviation of the input Gaussian distribution: slope $v$ and midpoint values $\mu$ determined as half point in the response function.

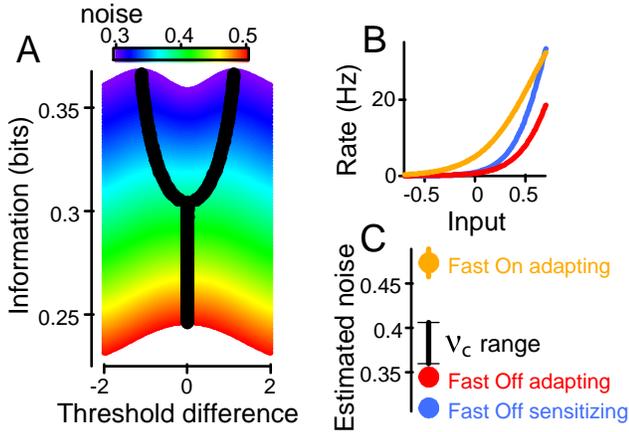

**Figure 2. Second-order phase transition accounts for differences in homogeneity of On and Off ganglion cell populations.** (**A**) In the model, information in shown as a function of the threshold difference, $m$, between midpoints of the two response functions with the same slope, $v$, representing the average noise level and denoted by color. In all cases shown here the rate constraint, $\langle r \rangle$, was the same. Black line shows the location of the maximal information for a given slope value. Bifurcation indicates the transition in the optimal solution from one to two populations. (**B**) On cells have shallower slopes than fast Off cells. Example nonlinearities for a fast Off sensitizing, a fast Off adapting, and an On cell recorded simultaneously. Input values are normalized such that their standard deviation (s.d.) is equal to 1. (**C**) Average slopes for fast Off sensitizing (n = 95), fast Off adapting (n = 388), and On (n = 58) cells. The critical point $v_c$ depends on the average rate $\langle r \rangle$, and is shown for the range of rates found in the data. For each $\langle r \rangle$, $v_c$ was determined by fitting the equation $m \propto |v - v_c|^\beta$ to the dependence of optimal threshold difference $m$ on noise $v$, from panel A, see also Figure 4C.

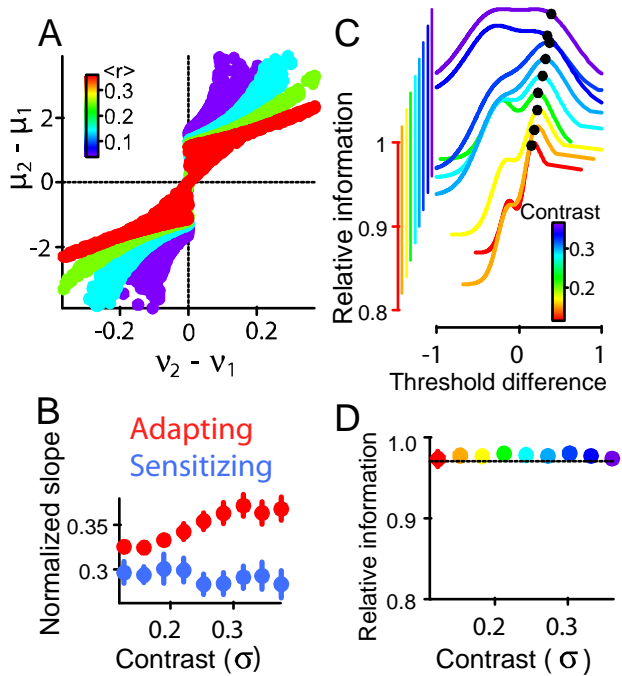

**Figure 3. Optimal dynamic range placement by fast Off populations.** (**A**) Optimal threshold difference, $m = \mu_2 - \mu_1$, as a function of the differences in the slopes, $v_2 - v_1$, and the average rate $\langle r \rangle$ (color), which is constrained to a fixed value for each optimal solution. (**B**) Average slope values normalized by the contrast, $\sigma$, for the same set of adapting and sensitizing cells from Figure 1B. (**C**) Information as a function of threshold difference between two response functions with slopes and a rate constraint taken from simultaneously recorded adapting and sensitizing cells. The black dots show the measured threshold difference at each contrast. Each curve was normalized by the maximum information at that contrast (denoted by color). The curves are vertically offset from each for better visualization. (**D**) The average percentage of the maximum information reached for all cells pairs (n = 7) at each contrast is > 97% (dotted line). Colors correspond to the colors from c. Error values, s.e.m., are obscured by the data points.

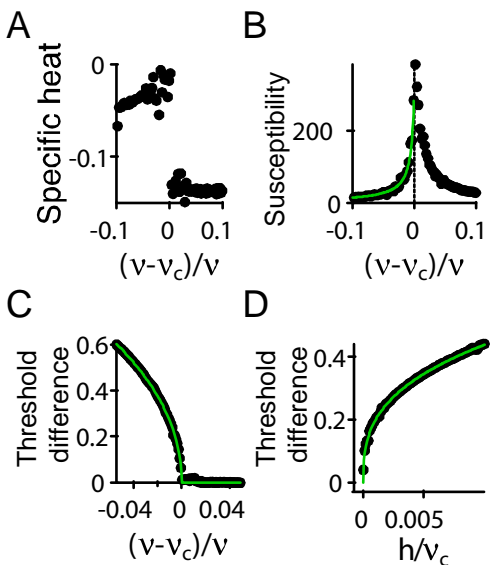

**Figure 4. Singular behavior of maximally informative solutions near the critical point matches the Ising model.** (**A**) The analogue of the specific heat, the second derivative of the information, $I$, with

respect to the noise, $C = \partial^2 I / \partial v^2$, exhibits a drop expected from mean-field calculations for the Ising model for $v > v_c$ [18, 19] (**B**) The analogue of the magnetic susceptibility, the second derivative $\chi = \partial^2 I / \partial h^2$ of the information, $I$, with respect $h = v_2 - v_1$, the quantity analogous to magnetic field, diverges as $|v - v_c|^{-1}$. The fit (green line) yields an exponent of -0.93, which matches the value of -1 predicted by the mean-field theory for the magnetic susceptibility. (**C**) The optimal threshold difference between response functions as a function of the slope, $v$, follows the theoretically predicted equation $m \propto \sqrt{|v - v_c|}$ dependence, when $h=0$. The fit of the relationship yields $m \propto |v - v_c|^{0.47}$ (green line). (**D**) Optimal threshold difference as a function of the slope difference, $h$, a quantity analogous to a magnetic field, for $v = v_c$ follows the predicted dependence of $m \propto h^{1/3}$. The fit of the relationship yields $m \propto h^{0.34}$ (green line). The functions in A-C are plotted relative to $(v - v_c)/v$, the normalized distance to the critical point, such that zero indicates that $v$ is at the critical noise value. This difference is normalized by $v$ following definition from [29] to obtain a dimensionless quantity, analogous to reduced temperature.

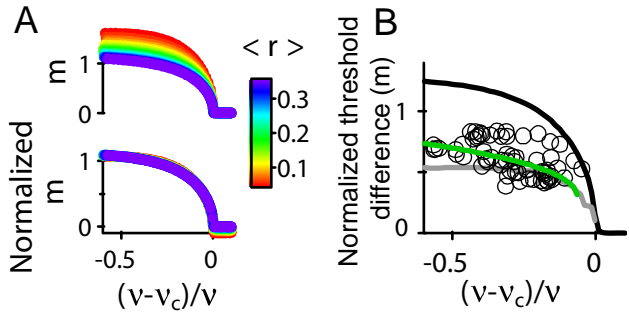

**Figure 5. Combined data from pairs with different average spiking probabilities shows that scaling exponents in the retina match the Ising model universality class.** (**A**) In the model, the threshold difference $m$ between cells is plotted against the normalized critical temperature, $(v - v_c)/v$ without normalizing for the average rate $\langle r \rangle$ (top), and with normalization (bottom). (**B**) Threshold difference observed experimentally normalized by threshold difference that maximized the information (black dots) and the spinodal line (grey curve), which delineates the regions between fast and slow dynamics near a critical point. Green line shows the average fit for the equation used to determine the scaling exponents (Table 1), using the average $h$ value found in the data. All exponents take similar values whether the reduced effective temperature is defined as $|v - v_c|/v$ or $|v - v_c|/v_c$ (not shown). The first definition is in according with notations from[29].

**Table 1. Mapping between maximally informative solutions in neural circuits and the canonical Ising model of phase transitions in physics.**

|  | Magnetic systems (Ising model) | Maximally informative coding |
|---|---|---|
| **Optimal states defined by:** | Minima of free energy | Maxima of information |
| **Transitions occurs with respect to:** | Temperature | Input noise (average slopes of neural response functions) |
| **Symmetry broken below the critical temperature** | Magnetization direction | Exchange symmetry between neurons |
| **Order parameter** | Magnetization | Deviation of thresholds from the mean (difference for n=2) across a neural population* |
| **Conjugate field** | Applied magnetic field | Deviation of slopes from the mean (difference for n=2) across a neural population |
| **Exponent with respect to temperature for $h=0$** | Mean-field value: ½<br>Experiment: 0.316 – 0.327 | Mean-field value: ½<br>Our experimental value: $0.39 \pm 0.12$ |
| **Critical isotherm exponent** | Mean-field value: ⅓<br>Experiment: 0.2 – 0.21 | Mean-field value: ⅓<br>Our experimental value: $0.15 \pm 0.08$ |

The order parameter is the parameter that measures the degree to which solutions below the critical point deviate from the symmetric solution present above the critical point. The exponent $\beta$ comes from the fit to the relationship between the threshold difference and the noise (Figure 4C). The exponent $1/\delta$ comes from the fit to the relationship between the threshold difference and $h$, the difference between slopes of the two response function (Figure 4D). The values (± s.d.) for the scaling exponents ($\beta$ and $1/\delta$) for our system were determined by a Bootstrap fit to the 7 pairs of adapting and sensitizing cells in the 9 different contrasts, using the equation $A|v - v_c h|^{\beta} + Bh^{1/\delta}$ (Figure 5B). * A quantitative verification of this definition is provided in Figure S3.

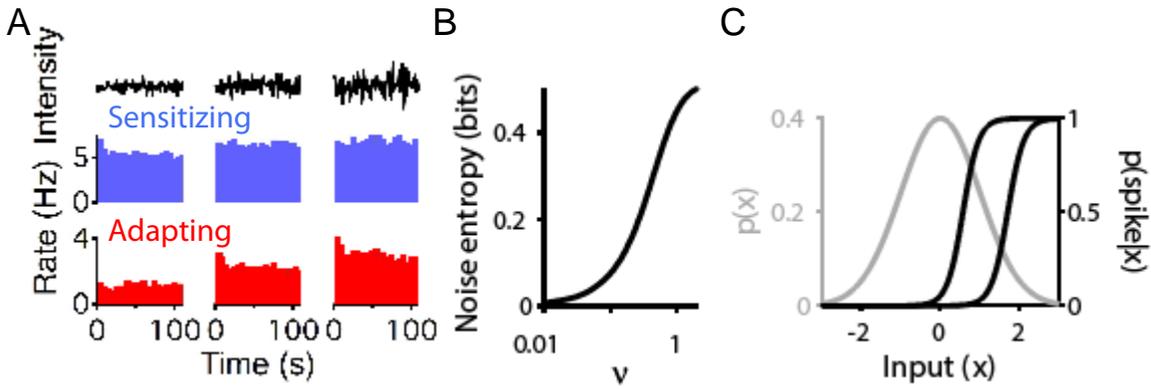

**Figure S1. Experimental paradigm and model exploration.** (A) Example neural responses for different levels of contrast. Stimulus (top) and average response of a fast Off sensitizing (middle) and adapting (bottom) cell recorded simultaneously. Cells are the same as in Figure 1A. (B) The noise entropy, evaluated according to Eq. (4), increases with increasing slope, $v$, for a single response function and a fixed value of $\langle r \rangle$. (C) An example of the maximally informative placement of two response functions for a given input distribution (gray line) and average response rate $\langle r \rangle$ shows the separation of thresholds between the two cell types.

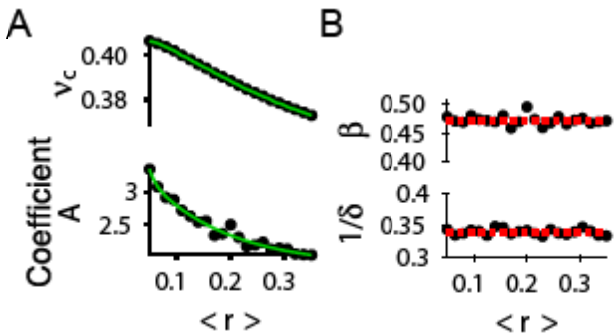

**Figure S2. Changes in model parameters with changes in the rate constraint.** (A) Model parameters other than the universal scaling exponents, $\beta$ and $1/\delta$, vary smoothly with $\langle r \rangle$. Top: the relationship between the critical noise value $v_c$ and $\langle r \rangle$. Bottom: the relationship between the proportionality coefficient $A$, from the fit to the model for the equation $m = A|v - v_c|^\beta$ when $h = 0$ (see Figure 4C), and different $\langle r \rangle$. (B) The scaling exponents $\beta$ (top) and $1/\delta$ (bottom) do not vary with the average rate, $\langle r \rangle$. Dotted red lines show the average values for the two exponents (0.47 and 0.34, respectively).

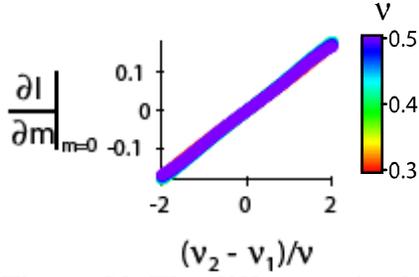

**Figure S3. The difference in slopes of neural response functions is analogous to a magnetic (conjugate) field.** In physics, a magnetic field induces a linear change in the average magnetization regardless of temperature. More specifically, in the Ising model, the value of the magnetic field can be found by taking the derivative of free energy with respect to magnetization at one of the optima. Performing an analogous procedure in the neural context amounts to evaluating the derivative of information with respect to $m$, for $m=0$. This yields a function that is proportional to the difference of the slopes of the two response functions $v_2 - v_1$, confirming that the latter quantity can be used as a proxy for the magnetic field. The analysis was repeated for multiple values of $v$ in different colors. The lines overlay, thereby obscuring the different colors.

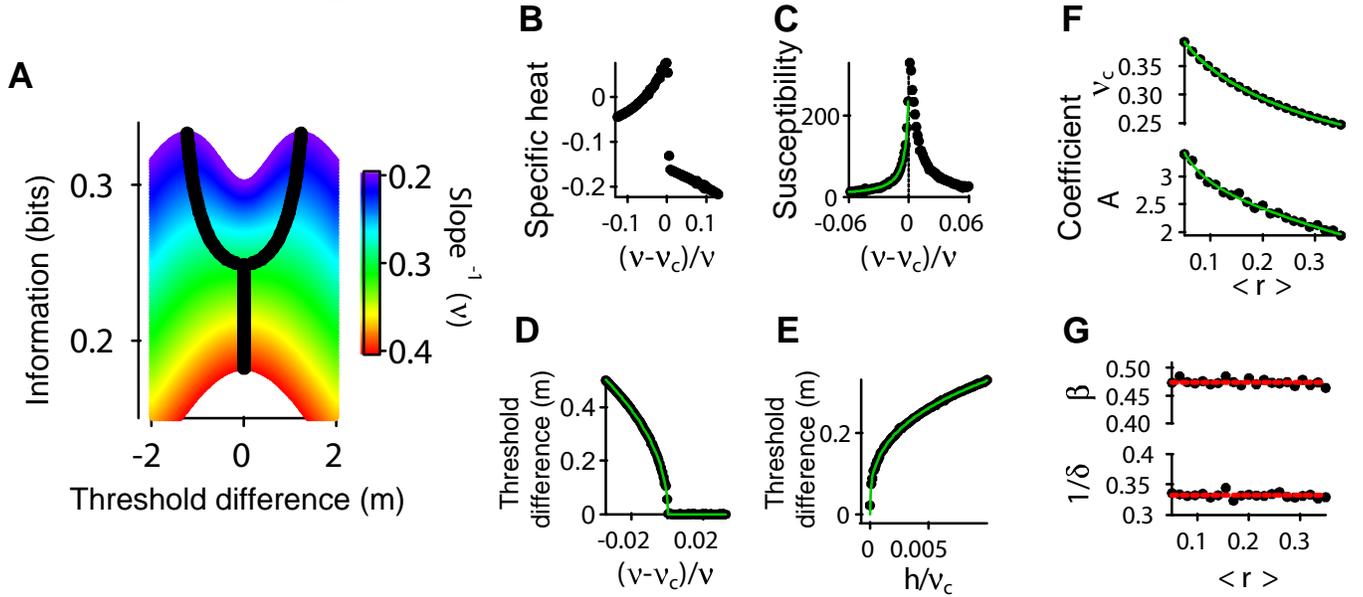

**Figure S4. Analysis of the critical point for non-Gaussian inputs.** The quantitative characteristics of the critical point are preserved when non-Gaussian inputs $P(x) \propto e^{-|x|}$ were used instead of a Gaussian distribution. (A) analogous to Figure 2A; (B)- (E) analogous to Figure 4, and (F)-(G) to Figure S2.

**Table S1.** Summary of predictions for critical exponents for nearest-neighbor Ising model in different dimensions. Data represents compilation of Table 3.1 from (Goldenfeld ,1992) and Table 3.4 from (Stanley, 1971) in comparison to our measurements.

| Exponent | 2D Ising | 3D Ising | 4D Ising/mean field | Experiments in various physical systems | Our measurements in the retina |
|---|---|---|---|---|---|
| β | 1/8=0.125 | 0.325±0.0015 | ½=0.5 | 0.316-0.35 | 0.39 ± 0.12 |
| δ | 15 | 4.82(4) | 3 | 4.2-4.9 | 7±4 |